\documentclass{ismdproc}

\begin{document}
\title{Studies of jet production with CMS in pp collisions at $\sqrt{s}$=7 TeV}


%

%

%
\author{{\slshape Panagiotis Katsas}  for the CMS Collaboration\\[1ex]
Deutsches Elektronen-Synchrotron (DESY), Notkestra\ss e 85, 22607 Hamburg, Germany }

\contribID{xy}  
\confID{yz}
\acronym{ISMD2010}
\doi            

\maketitle

\begin{abstract}
Preliminary results on a variety of jet analyses are presented using pp
collision data collected by the CMS experiment at $\sqrt{s}$=7 TeV.  We report
on measurements of the inclusive jet cross sections, the ratio of the
inclusive three-jet over two-jet cross sections, the hadronic event shapes,
the jet structure and the dijet azimuthal decorrelations. The data distributions are
compared with the predictions of Monte Carlo event generators and with
perturbative QCD calculations.
\end{abstract}

\section{Introduction}
Jets are the experimental signatures of quarks and gluons which manifest themselves as collimated streams of particles. At the LHC, the large cross section for jet production
allows to perform early studies of new kinematic regimes, hence confronting the predictions of perturbative QCD and probing physics processes within or beyond the standard model. A detailed description of the CMS detector can be found in \cite{cms}. CMS uses a right-handed coordinate system with z-axis parallel to the LHC beam direction, y-axis perpendicular to it and pointing upwards, azimuthal angle $\phi$ and polar angle $\theta$. A superconducting solenoid of 6 m internal diameter constitutes the main feature of the detector, providing a uniform magnetic field of 3.8 T. The inner tracking system is composed of a pixel detector with three barrel layers at radii between 4.4 and 10.2 cm and a silicon strip tracker with 10 barrel detection layers extending outwards to a radius of 1.1 m. Each system is completed by two endcaps, extending the tracking detector acceptance up to a pseudorapidity $|\eta|$=2.5. The calorimeters inside the magnetic coil consist of a lead-tungstate crystal electromagnetic calorimeter (ECAL) and a brass-scintillator hadronic calorimeter (HCAL). They provide a hermetic coverage over a large range of pseudorapidity ($|\eta| <$3 for ECAL and $|\eta| <$5 for HCAL). The calorimeter cells are grouped in projective towers of granularity $\Delta\eta \times \Delta\phi$ = 0.087 $\times$ 0.087 at central rapidities. 

\section{Jet reconstruction}
Three different types of jet reconstruction are employed by CMS, used as input to the anti-k$_T$ clustering algorithm \cite{akalgo} with distance parameter R=0.5. Calorimeter jets are reconstructed using energy deposits in the electromagnetic and hadronic calorimeters. The Jet-Plus-Tracks (JPT) algorithm \cite{jpt} is track-based and provides a correction to the energy and the direction of calorimeter jets exploiting the good performance of the CMS tracking detectors. The Particle Flow (PF) algorithm \cite{pflow} aims to reconstruct, identify and calibrate each individual particle in the event by combining information from every subdetector. The measured jet energy is corrected on average, from detector level to hadron level, according to a factorized approach consisting of different levels of corrections, which are applied sequentially\cite{JME-10-003}. A relative correction removes variations in the jet response as a function of the pseudorapidity, by calibrating with respect to the barrel region ($|\eta|<$1.3). An absolute correction ensures, furthermore, that the jet response does not depend on the transverse momentum. The jet energy scale uncertainties were estimated to be 10\% (5\%) for calorimeter (JPT and PF) jets. An additional 2\% uncertainty per unit rapidity was estimated from the relative correction.

\section{Trigger requirements and event selection}
CMS uses a two-tiered trigger system to select events online: a hardware Level-1 (L1) trigger and an online software High Level Trigger (HLT). Jets with relatively high transverse momentum, $p_T$, are recorded using single jet triggers, which require an L1 jet with uncorrected $p_T$ above 6 (20) GeV and an HLT jet with $p_T$ above 15 (30) GeV. Jets with lower transverse momentum are recorded with a prescaled Minimum Bias trigger, which required activity in both beam scintillator counters located at 3.23$<|\eta|<$4.65 in coincidence with colliding proton bunches. All triggered events are required to have a good primary vertex consistent with the measured transverse position of the beam. At least one vertex must be reconstructed in the event and the z-coordinate of the primary vertex is required to be within the luminous region, with $|z|<$15 cm. Additionally, its radial distance must be less than 0.15 cm and the fit for the primary vertex sufficiently constrained with at least five associated tracks. With these requirements, non-collision and beam related backgrounds are rejected. Loose quality criteria are applied to each jet, as described in \cite{caloID} and \cite{pfID}, depending on the type of jet.

\section{Inclusive jet cross section and 3-jet to 2-jet cross section ratio}
\begin{figure}[htb]
\vspace{0.6cm}
\hspace*{0.6cm}
\includegraphics[width=5.cm, height=3.2cm,angle=-90]{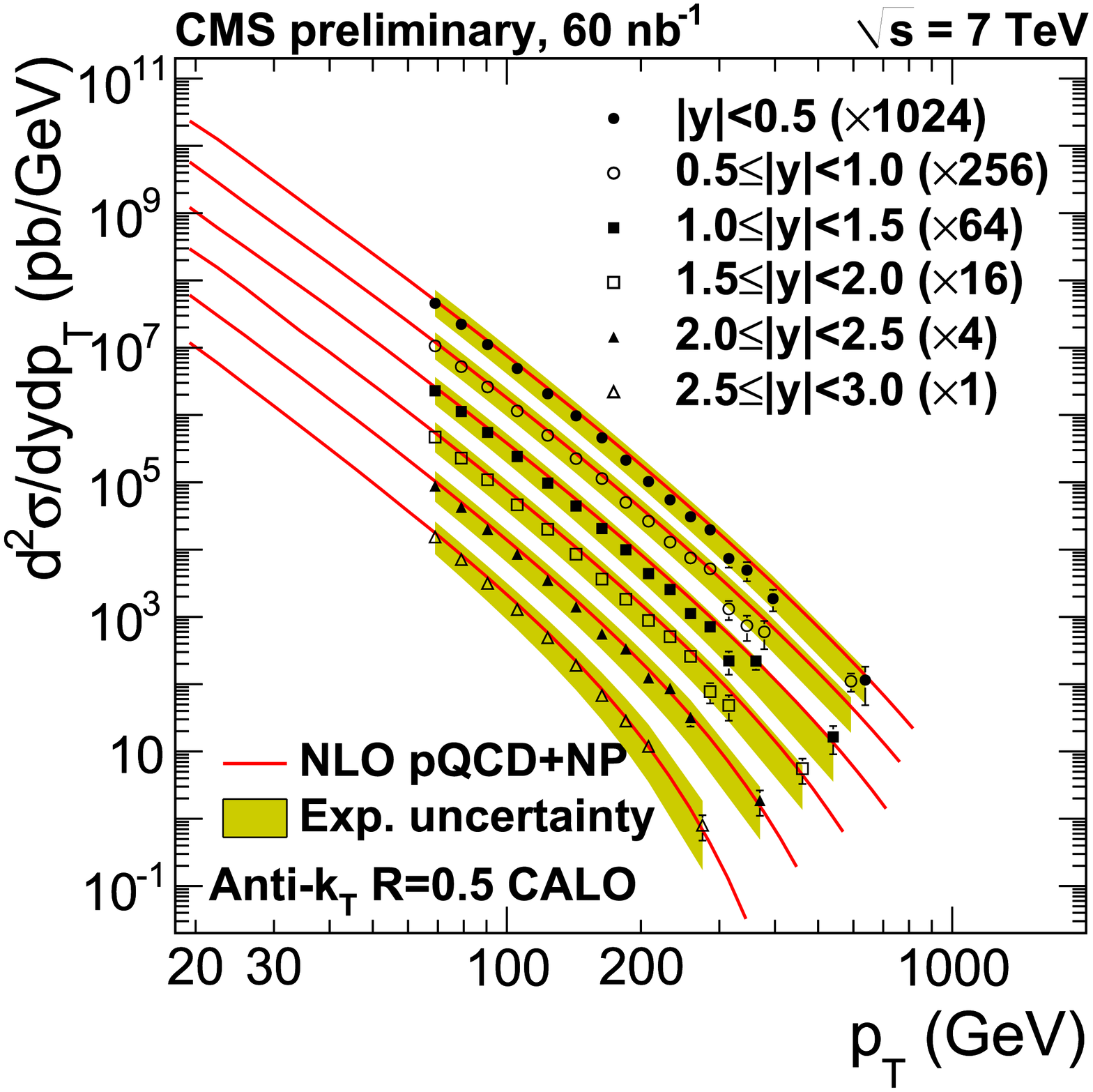}\hspace*{1.5cm}
\includegraphics[width=5.cm, height=3.2cm,angle=-90]{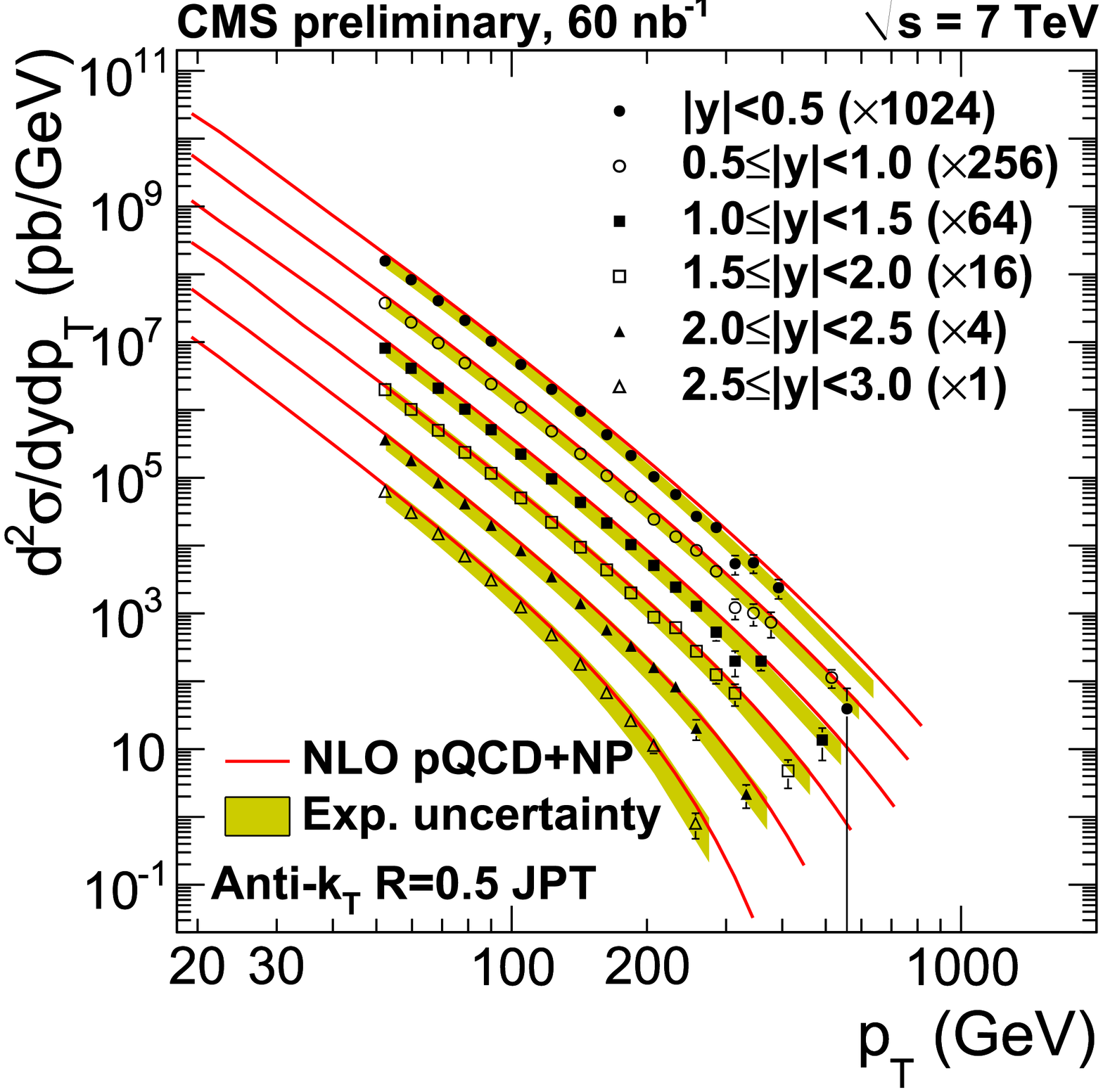}\hspace*{1.5cm}
\includegraphics[width=5.cm, height=3.2cm,angle=-90]{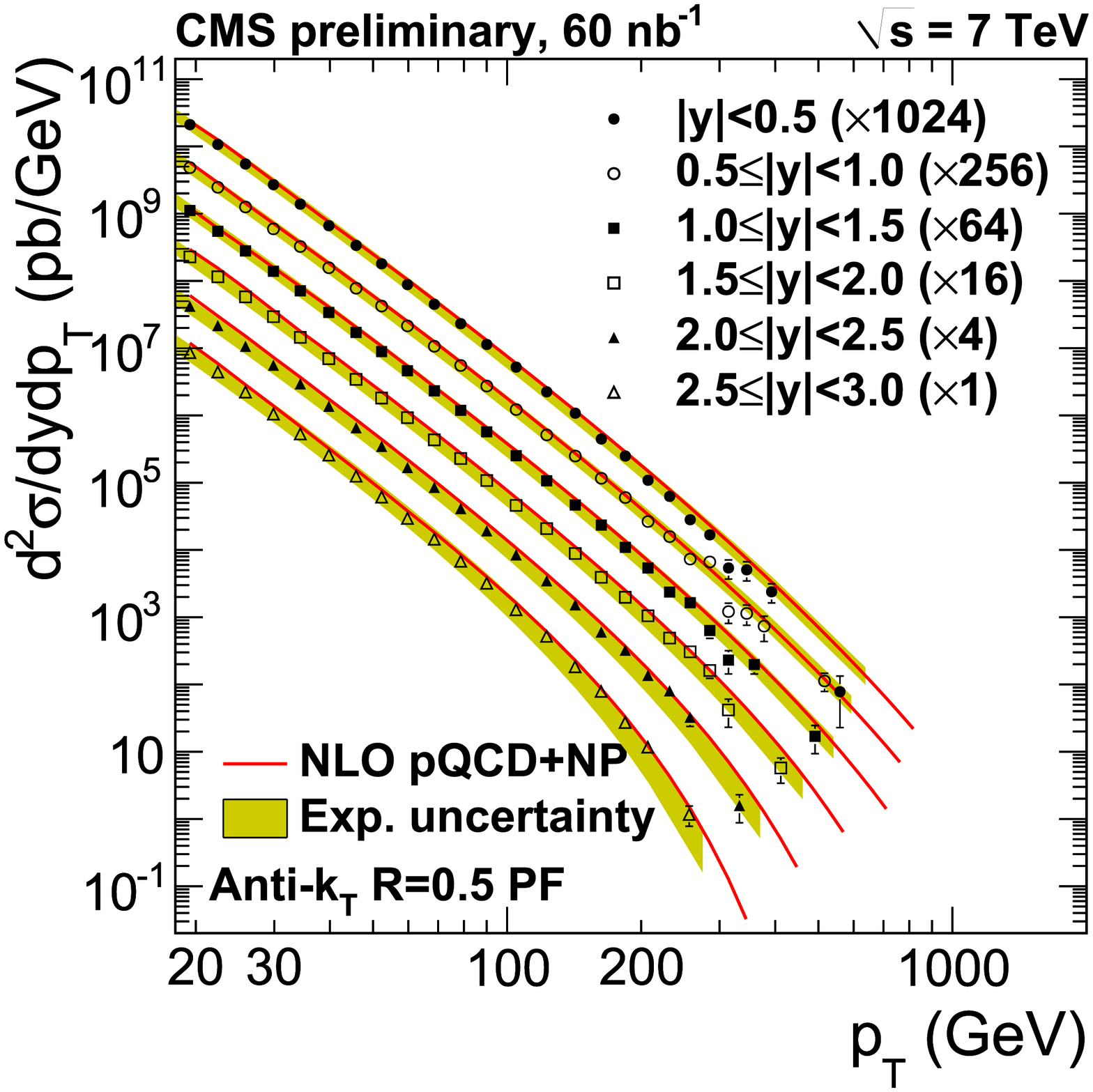}\vspace*{0.7cm}
\caption{Measured cross section as a function of the jet transverse momentum, in different bins of rapidity for calorimeter jets (left), Jet-Plus-Track jets (center) and Particle Flow jets (right). The solid line represents NLO predictions. For better visibility the spectra are multiplied by arbitrary factors, as indicated in the legend.}\label{Fig:fig1}
\end{figure}
The inclusive jet production cross section is one of the basic measurements performed
at hadron colliders. It is defined as \cite{qcd-10-011}:
\begin{equation}\label{eq:xsection}
\frac{d^2\sigma}{dp_Tdy} = \frac{C_{res}}{L \cdot \epsilon} \times \frac{N_{jets}}{\Delta y\cdot\Delta p_T}
\end{equation}
where $N_{jets}$ is the number of jets, $C_{res}$ is a correction factor for bin-to-bin migrations due to resolution effects, $\Delta p_T$ and $\Delta y$ are the bin widths in transverse momentum and rapidity, respectively, $L$ is the total integrated luminosity and $\epsilon$ is the product of event and jet efficiencies, as calculated per jet.
The spectrum of the transverse momentum of the jets is corrected for resolution effects using an ansatz. It is assumed that the true spectrum is modeled by a function of the jet transverse momentum, $f(p_T)$, which is based on early phenomenological fits partly motivated by the parton model \cite{bjorken1971},\cite{feynman1978}:
\begin{equation}
f(p_T) = N p_T^{-\alpha}\Big(1-\mathrm{cosh}(y_{min})\frac{2p_T}{\sqrt{s}}\Big)^{\beta}e^{-\gamma/p_T}
\end{equation}
The true spectrum is then smeared using known jet resolutions:
\begin{equation}
F(p_T) = \int_0^{\infty}f(p'_T)\Re(p'_T-p_T, \sigma)dp'_T
\end{equation}

\begin{figure}[htb]
\centerline{\includegraphics[width=8.2cm, height=5.cm,angle=0]{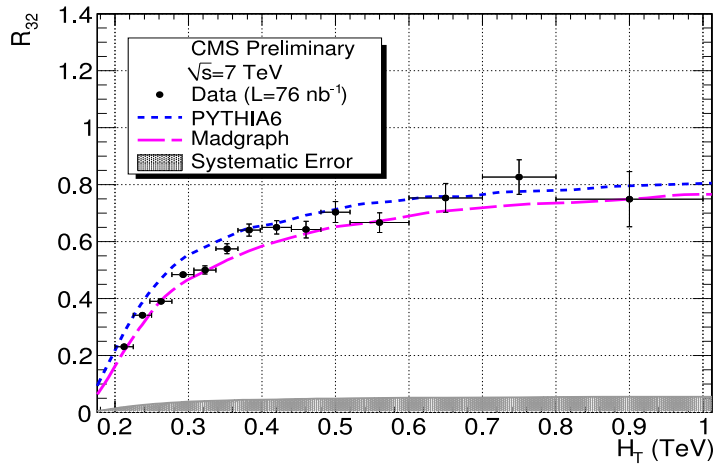}}
\vspace*{0.5cm}
\caption{Measured 3-jet to 2-jet cross section ratio as a function of the total jet transverse momentum. For comparison, \textsc{pythia} (dashed line) and Madgraph (solid line) predictions are shown. The shaded area indicates the systematic uncertainty.}\label{Fig:fig2}
\end{figure}

The smearing function $\Re(p'_T-p_T, \sigma)$ is assumed to be a Gaussian. The parameters of the model are determined by fitting the smeared transverse momentum spectrum, $F(p_T)$, to the data. The final corrections, inserted in Eq. (1), are determined from the ratio $C_{res}=f(p_T)/F(p_T)$. The measured cross section is shown in Figure \ref{Fig:fig1}, as a function of the transverse momentum, for different bins of rapidity and for different types of jets. The results agree with the theoretical prediction and with each other to within 20\%, over most of the measured $p_T$ and rapidity ranges.
The ratio of the inclusive 3-jet to 2-jet cross sections, $R_{32}$, was also measured as a function of the total scalar transverse momentum sum, $H_T$, for jets with $p_T>$50 GeV and $|y|<$ 2.5 \cite{qcd-10-012}. As shown in Figure 2, the ratio rises with $H_T$ as phase space opens up for a third jet to be emitted. The plateau, observed at about $R_{32}=0.8$, is sensitive to $\alpha_s$, however its exact value depends on the applied selection criteria and the jet finding algorithm.

\section{Jet transverse structure and momentum distribution}
\begin{figure}[htb]
\centerline{\includegraphics[width=6.cm, height=14.5cm,angle=-90]{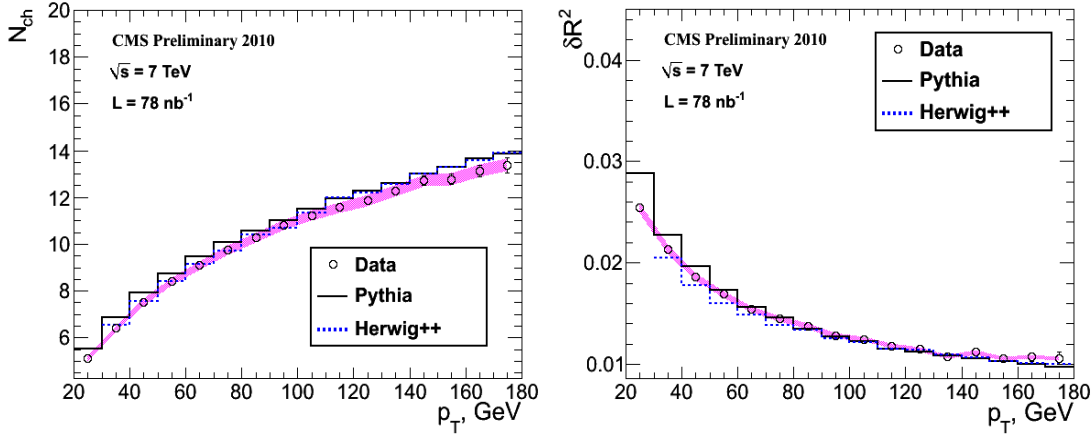}}
\caption{Charged particle multiplicity, $N_{ch}$ (left) and transverse jet shape $\delta R^2$ (right) as function of the corrected transverse momentum $p_T$ of Jet-Plus-Track jets. Data are shown with statistical error bars, with a band denoting the systematic errors. Predictions based on the \textsc{pythia} 6 and \textsc{herwig++} event generators are also shown.}\label{Fig:fig3}
\end{figure}

The structure of jets was studied \cite{qcd-10-014} by measuring their charged particle multiplicity, $N_{ch}$, and the charged particle transverse jet shape, which defines the width of a jet in the $\eta$-$\phi$ plane. The latter is given by $\delta R^2 = \langle \delta\phi^2\rangle + \langle \delta \eta^2 \rangle$, 
where the averages are second moments, using the transverse momenta of charged particles associated to a jet by the JPT algorithm. Figure \ref{Fig:fig3} depicts the measured averages of $N_{ch}$ and $\delta R^2$ as a function of the transverse momentum. The data are compared with predictions obtained from simulations with the \textsc{pythia} (D6T tune) and \textsc{herwig++} event generators and it is found that they are consistent within the quoted experimental uncertainties. At low $p_T$, the data suggest a few percent broader (narrower) jets than what is predicted by \textsc{herwig++} (\textsc{pythia}).

\section{Hadronic event shapes}
\begin{figure}[htb]
\vspace{0.4cm}
\centerline{\includegraphics[width=5.5cm, height=14.0cm,angle=-90]{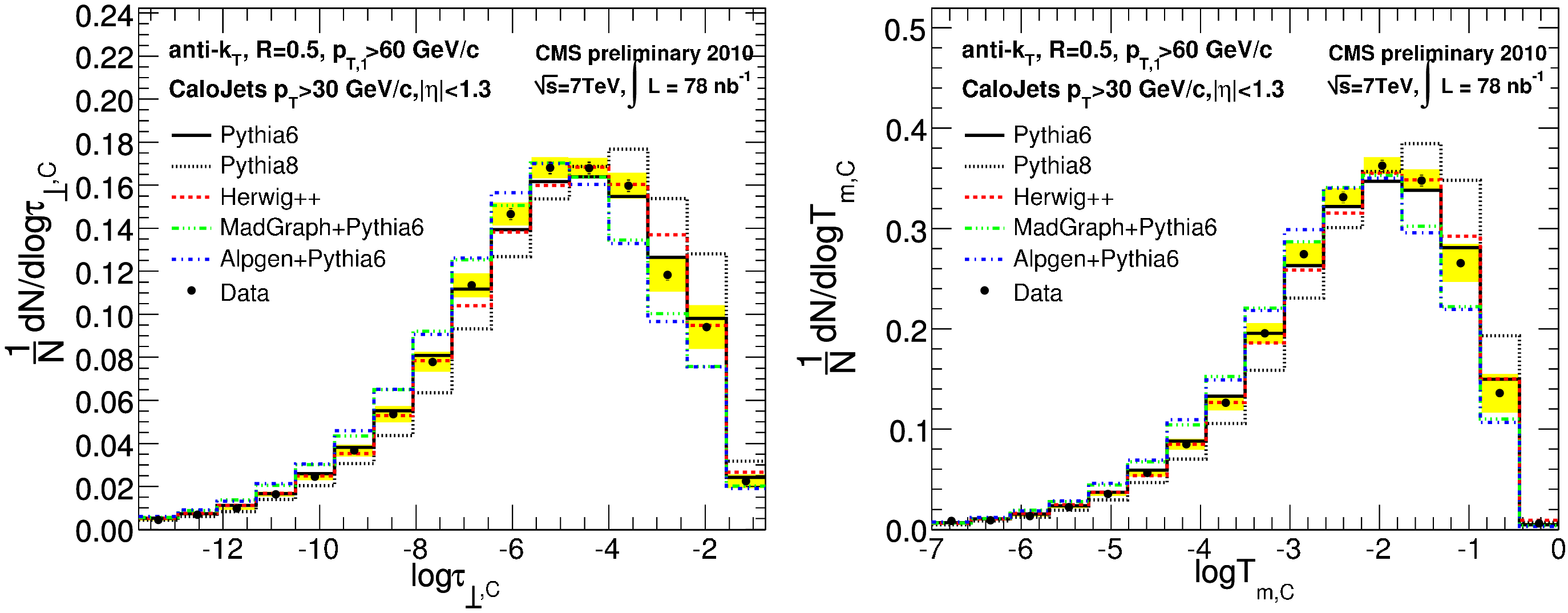}}
\vspace*{0.2cm}
\caption{The central transverse thrust (left) and central thrust minor (right) distributions for calorimeter jets in events with leading jet $p_T >$ 60 GeV/c.The yellow bands represent the sum of statistical and systematic errors.}\label{Fig:fig4}
\end{figure}

The hadronic event shapes provide geometric information about the energy flow in hadronic
events which can be exploited for the tuning of the parton shower and non-perturbative components of MC event generators. Measurements were performed focusing on two event shape variables: the central transverse thrust $T_{\bot,C}$ and the central thrust minor $T_{m,C}$ \cite{qcd-10-013}. As shown in Figure \ref{Fig:fig4}, the event shape distributions from \textsc{pythia} 6 and \textsc{herwig}++ show satisfactory agreement with the data, while discrepancies are found between the data and predictions from \textsc{alpgen}, MadGraph, and \textsc{pythia} 8.

\section{Dijet azimuthal decorrelations}
\begin{figure}[htb]
\vspace*{1.cm}
\centerline{\includegraphics[width=6.5cm, height=7.5cm]{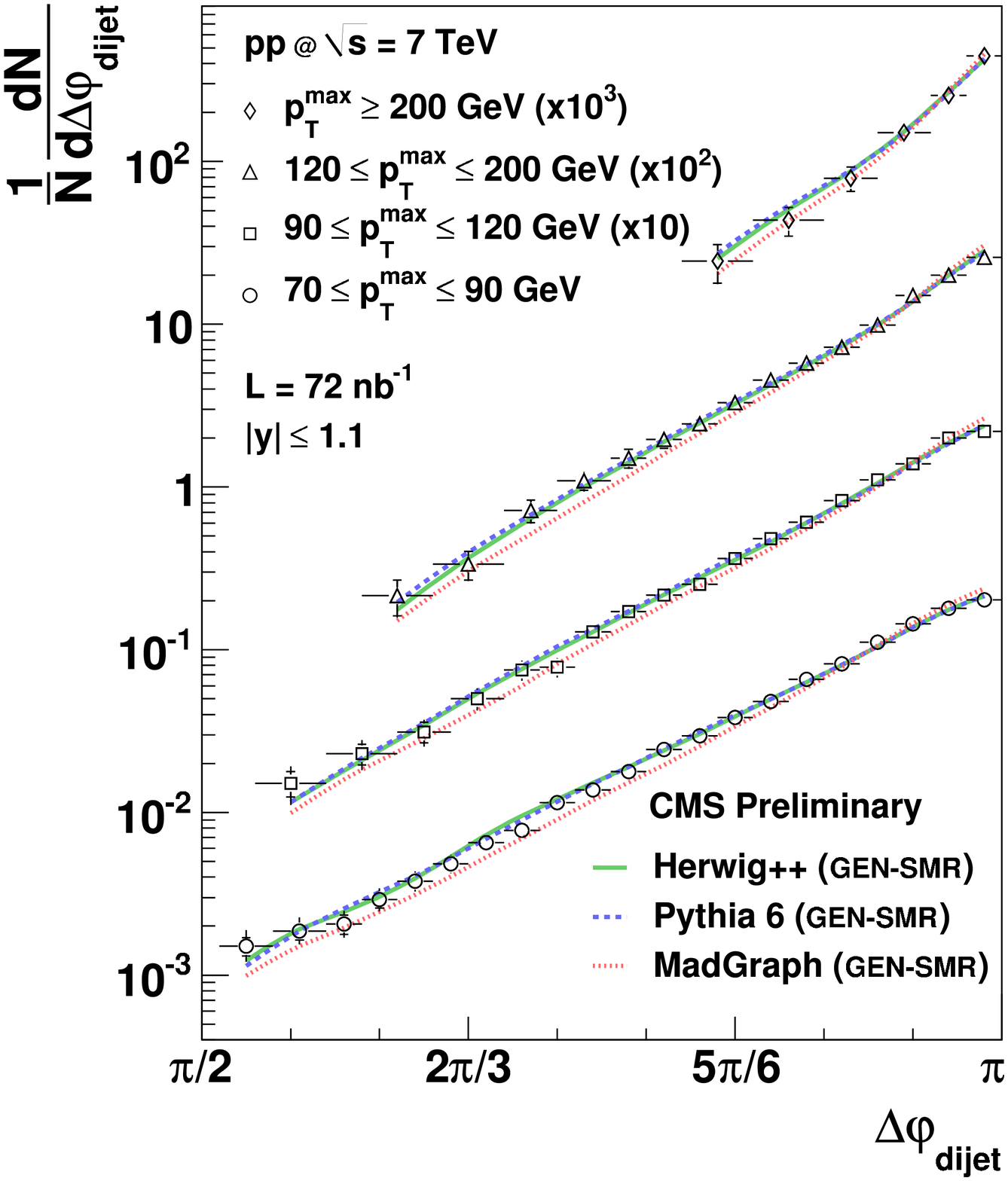}\vspace*{1.cm}
\includegraphics[width=6.5cm, height=7.5cm]{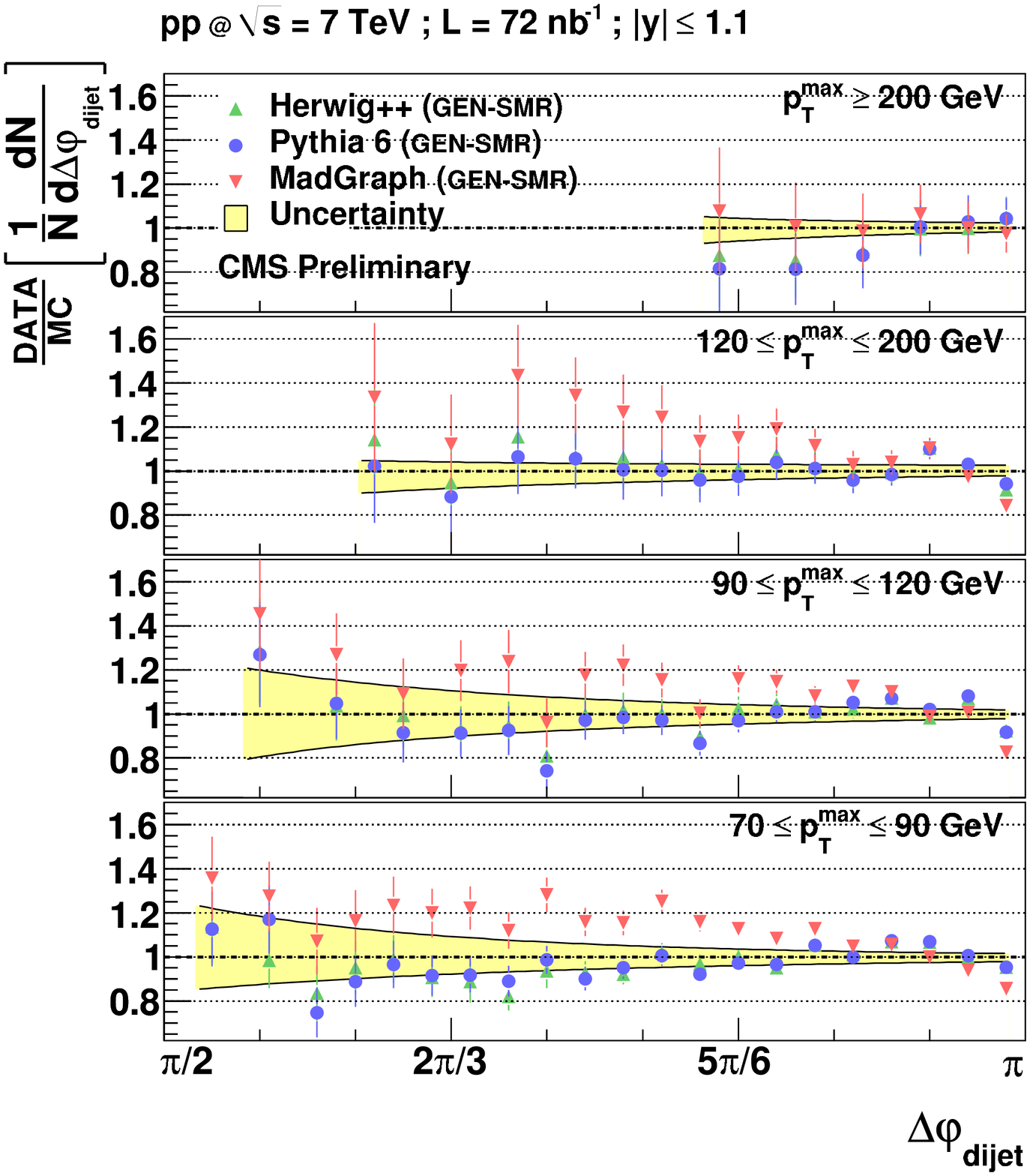}}
\vspace*{-1.cm}
\caption{Left: Event-normalized differential dijet distributions in $\Delta\phi_{dijet}$, for different leading jet $p^{max}_T$ ranges. Predictions from \textsc{pythia, herwig} and \textsc{madgraph} are also shown for comparison. Right: Ratios of the measured $\Delta\phi_{dijet}$ distributions to the ones predicted from simulations with \textsc{pythia, herwig} and \textsc{madgraph} in different leading jet $p_T$ regions. Detector resolution effects on jet $p_T$ and position have been included in the MC predictions at the generated particle level (GEN-SMR).}\label{Fig:fig5}
\end{figure}

The azimuthal angle, $\Delta\phi_{dijet}$, between the two highest transverse momentum jets in hard-scattering events can be used to study higher-order QCD radiation effects without the need to to explicitly reconstruct additional jets. The normalized differential distributions in $\Delta\phi_{dijet}$ for the data and several MC event generators are shown in Figure \ref{Fig:fig5}.
The sensitivity of the distributions to QCD initial-state radiation and final-state
radiation effects was investigated by varying the multiplicative parameters that control their amount in \textsc{pythia}. It was found that they are most sensitive to the initial-state radiation, while varying the amount of final-state radiation causes very small effects\cite{qcd-10-015}.

\section{Acknowledgments}
The author would like to thank the organizers of the conference for their hospitality and Nikos Varelas for useful comments while preparing the proceedings.

\begin{footnotesize}

\end{footnotesize}


\end{document}